\documentstyle[12pt]{article}

\def\beq{\begin{equation}}
\def\eeq{\end{equation}}

\def\bea{\begin{eqnarray}}
\def\eea{\end{eqnarray}}
\def\ben{\begin{enumerate}}
\def\een{\end{enumerate}}

\def\ni{\noindent}
\def\nn{\nonumber}

\def\ms{\medskip}

\def\tr{\mbox{Tr}\, }
\def\brr{\begin{array}}
\def\err{\end{array}}

\begin{document}




\begin{center}

{\LARGE \bf
Bosonic string with antisymmetric fields and a non-local
Casimir effect}

\vspace{8mm}

{\sc E. Elizalde}\footnote{E-mail: eli@zeta.ecm.ub.es} \\
Center for Advanced Studies CEAB, CSIC, Cam\'{\i} de Santa
B\`arbara,
17300 Blanes,
\\ and Department ECM and IFAE, Faculty of Physics,
University of Barcelona, \\ Diagonal 647, 08028 Barcelona,
Catalonia, Spain \\
 and \\
{\sc S.D. Odintsov}\footnote{E-mail: odintsov@ecm.ub.es.}
\\
 Tomsk Pedagogical Institute, 634041 Tomsk, Russia, \\
and Department ECM, Faculty of Physics,
University of  Barcelona, \\  Diagonal 647, 08028 Barcelona,
Catalonia,
Spain \\

\vspace{2cm}

{\bf Abstract}
\end{center}

The coupling, in a non-standard way, of a bosonic string theory
with a
dilaton and antisymmetric fields is investigated. By
integrating over
the antisymmetric fields, a Coulomb-like interaction term
is generated. The
static potential of a theory of this kind is obtained from the
corresponding non-local zeta function, in some
approximation. An interpretation of the static potential as a
type of non-local Casimir effect is given.

\vfill

\newpage

\ni{\bf 1. \ Introduction.}
It escapes to nobody nowadays that, in spite of its long history, string
theory
(see \cite{2} for a general review) is not a completely understood
discipline yet.
This is reflected, in particular, by the fact that different versions,
formulations and
modifications of string theories are still being actively
studied.
Among the most interesting, recent modifications to this theory
one
should include the cases of a string coupled to a background
field
\cite{1}, the rigid string \cite{8}, and the Dirichlet string
theory
\cite{13}. Some months ago, the coupling of a rigid string to
antisymmetric fields was discussed, Ref. \cite{4}. This
formulation
is certainly different from the by now standard case of a string
in a
background field \cite{1}, because in Ref. \cite{4} the kinetic term
for the
antisymmetric fields had to be added by hand. Later, the
antisymmetric
fields can be integrated out, what leads to a Coulomb-like
term in the potential.

In the present paper we shall consider the usual bosonic string
theory,
interacting, in the  above way, both with antisymmetric and
dilaton
fields. The precise form of the interaction of Coulomb type that
appears after integrating out the background fields will be
found. We
shall see that the static potential in such a theory leads to a
very
non-standard zeta function, which can be interpreted as
originating a
non-local Casimir effect. A careful (although necessarily
approximate)
study of this static potential ---which arises from the corresponding
(non-local) zeta function--- will be given.
\ms

\ni{\bf 2. \ String theory coupled to a dilaton and
antisymmetric fields.}
Consider the action of a closed bosonic string in a field of
massless modes:
\beq
S= k \int d^2\xi \, \sqrt{g} \, \left[ \frac{1}{2} G_{ij}(X)
g^{\mu\nu} \partial_\mu X^i  \partial_\nu X^j + \zeta R^{(2)}
\phi (X) + \frac{e}{\sqrt{g}} \epsilon^{\mu\nu} A_{ij}
\partial_\mu X^i  \partial_\nu X^j\right],
\label{551}
\eeq
where $i=1, \ldots, D$ and $\mu =1,2$, $ G_{ij}$ is a symmetric
tensor (the source of the graviton modes), $k$ is the string
tension (which we will set equal to 1 in this section),  $A_{ij}$
is an antisymmetric tensor (the source for the antisymmetric
2-tensor modes), and $\phi (x)$ is the dilaton field (notice that
we do not consider the tachyon). In the standard approach to the
string effective action \cite{1,2}, the kinetic terms for the
sources that appear in  (\ref{551}) would show up explicitly
after integrating over the metrics.

However, our approach here will be different. Let us imagine that
string theory is coupled to some fields ---which can simply be
considered as external ones. In particular, we may consider a
theory  coupled to a dilaton and Kalb-Ramond fields
\cite{3}, in which case one has to add to the action (\ref{551})
the corresponding kinetic terms for those fields:
\beq
S_{kin}=  \int d^Dy \, \sqrt{g(y)} \, \left[ \frac{1}{2} \phi
\Box_D \phi  + \frac{1}{12} F_{ijk} F^{ijk}\right],
\label{552}
\eeq
where  $ F_{ijk} $ is the stress tensor for $A_{ij}$ and where
the kinetic terms in (\ref{552}) are defined in a $D$-dimensional
curved spacetime, being $\Box_D$ the d'Alembertian  in such
spacetime. As usually, $G_{ij}$ is a more fundamental object than
$A_{ij}$ and $\phi$, since it plays the role of a kinetic term
for the bosonic string. This is the reason why we do not need to add any
term of the form (\ref{552}) for it. Furthermore, there is no
possibility of guessing a simple kinetic term corrsponding to $G_{ij}$,
due to its geometrical structure.

The source terms in (\ref{551}) can be written as follows
\beq
 \int d^Dy \, \sqrt{g(y)} \, \left[ K(y) \phi  +
\frac{1}{\sqrt{g}} K^{ij} (y) A_{ij}\right],
\label{553}
\eeq
where
\bea
K(y)& =& \int d^2\xi \, \sqrt{g} \, \left[ \zeta R^{(2)}(\xi )
\delta (y- X(\xi ))\right], \nn \\
K^{ij}(y)& =& \int d^2\xi \,  \left[ e \,  \epsilon^{\mu\nu}
\partial_\mu y^i  \partial_\nu y^j \delta (y- X(\xi ))\right].
\label{554}
\eea
One can see that the functional integrals over
$A_{ij}$ and $\phi$ have the standard Gaussian form. Thus, we can
integrate over $A_{ij}$ and $\phi$ in order to obtain an
effective theory for the closed bosonic string, exhibiting a
Coulomb-like interaction term (compare with Ref. \cite{4}
and we also suppose that $D=4$ ,otherwise the form
of potential will be different)
\bea
S&=& \int d^2\xi \, \sqrt{g} \, \left[ \frac{1}{2} G_{ij} (X)
g^{\mu\nu} \partial_\mu X^i  \partial_\nu X^j \right] + S_{int},
\nn \\
S_{int}&=& \int d^2\xi d^2\xi' \, \left[ c_1 e^2 \sigma_{ij} (\xi
) \sigma^{ij} (\xi' ) + c_2 \zeta^2 {R^{(2)}}^2 \right] V(|x-
x'|),
\label{555}
\\ &&
\sigma_{ij} (\xi ) =  \epsilon^{\mu\nu}  \partial_\mu X^i
\partial_\nu X^j, \ \ \ \  V(|x-x'|) =\frac{1}{|x(\xi )- x(\xi'
)|^2+a^2}, \nn
\eea
being $c_1$ and $c_2$ non-essential numerical constants which can
be choosen to be equal to 1. It is the coupling of the
bosonic string with the kinetic terms of the form (\ref{552})
what induces the higher derivatives in the effective action and
the non-local interaction term. In what follows we shall
consider, for simplicity, the situation where the two-dimensional
space is flat, and hence $R^{(2)}=0$. Notice that in the
potential, $V$, a cut-off parameter, $a^2$, has been introduced,
in order to avoid the singularity that occurs at $x=x'$.
That is just usual ultraviolet-type cut-off.
The appearance of a term of the form of the one obtained here
---induced by the antisymmetric tensor field--- has been mentioned in
Ref. \cite{4} in connection with the rigid string theory.

An interesting question is now, to which consequences can this
Coulomb-like term lead in the frame of string theory. We are
going to discuss this question in some detail, by means of the
evaluation of the static potential for the model (\ref{555}) in
the case of a flat metric $G_{ij}= \delta_{ij}$.
\ms

\ni{\bf 3. \ The static potential, from a non-local zeta function.}
The static potential in string theory is an interesting magnitude
in connection with the possible applications of string theory to QCD.
This fact was realized long ago \cite{6}, and a calculation of
the static potential in different string models has been carried
out explicitly in Refs. \cite{5,7}, and in particular for the
rigid string \cite{8} in Refs. \cite{9}-\cite{11}. It has been
pointed out in those works that the leading corrections to
the static potential have a universal character \cite{5}.

In the applications of the static potential to QCD one can usually
choose the Wilson loop $C$ to be a rectangle on the plane, of
length $T$ and width $R$ (with $T>>R$). Then, the loop
expectation value can be found as follows \cite{12}:
\beq
W[C] \sim \exp [-TV_s (R)].
\label{556}
\eeq
The explicit one-loop calculation for a standard bosonic string
gives:
\beq
V_s(R) = kR + \frac{D-2}{2} \tr \ln \Box,
\label{557}
\eeq
where $k$ is again the string tension and $\Box$ the
two-dimensional d'Alembertian in the space of topology $R\times
S^1$. Using the zeta-function regularization procedure to
calculate (\ref{557}) one obtains the well-known result
\cite{5,7,enc}
\beq
V_s (R)= kR - \frac{(D-2)\pi}{24R}.
\label{558}
\eeq

We shall now perform the calculation to one-loop of the static
potential (\ref{557}) taking into account the Coulomb-like term
induced by the antisymmetric tensor fields, as in (\ref{555}).
We discover that the integration over the $X^i$s is not
Gaussian any more and this makes the calculation highly
non-trivial. We shall however prove that a meaningful result can
be obtained in a quite reasonable approximation.
The natural way to consider the integration is by decomposing the
variables as follows:
\beq
X^i(\xi) = X_0^i(\xi) + X_1^i(\xi),
\label{559}
\eeq
where $ X_0^i(\xi)$ is a background variable, which is a linear
function of $\xi$ satisfying the field equations
\beq
X_0^i (\xi) = c_\mu^i \xi^\mu, \qquad \eta_{ij} c_\mu^i c_\nu^j
=\eta_{\mu\nu},
\label{5510}
\eeq
being  the $c_\mu^i$ some  constants. The expansion of Eq.
(\ref{555}) up to second order on the fluctuations
$X_1^i(\xi)$ can be performed in the same way as it was
done in Ref. \cite{4}. After the subsequent functional Gaussian
integration over the $X_1^i$s, we obtain the static potential
(for simplicity we consider below only case $D=4$)
\beq
V_s(R)= kR + \frac{D-2}{2} \int dp
\sum_{n=1}^\infty \ln \left[ \frac{p^2}{2} + 2e^2p^2 \int d^2\xi \,
\frac{e^{ip\cdot \xi}}{\xi \cdot \xi +a^2}
 + 4e^2 \int d^2\xi \, \frac{e^{ip\cdot \xi}-1}{(\xi \cdot
\xi +a^2)^2} \right].
\label{5511}
\eeq
Notice that the notation has been simplified somehow, because in
(\ref{5511}) it must be understood that the double integration
over $d^2\xi$ is in fact a single integration on the first
coordinate $\xi_1$ and an infinite sum (one of the spatial
coordinates corresponds to the torus), exactly as in the case of
the first integration (over $p$ and $n$). For the benefit of the
reader, let us here briefly illustrate this case, where the procedure
of zeta-function regularization \cite{zfr1,dowk1} is self-explanatory
(for a very detailed review, see \cite{zb1})
\bea
&& \int d^2p \ (p\cdot p)^{-s} = \int_0^\infty dp \sum_{n=1}^\infty
\left( p^2 + \frac{n^2}{R^2} \right)^{-s}  \label{5512} \\  &&
\hspace{3mm} = \frac{1}{\Gamma (s)}  \int_0^\infty dt \, t^{s-1} e^{-
n^2 t/R^2} \int_0^\infty \sum_{n=1}^\infty dp \ e^{-p^2 t} = -
\frac{\sqrt{\pi} \Gamma (s-1/2) R^{2s-1}}{2\,\Gamma (s)} \, \zeta
(2s-1), \nn
\eea
and taking the derivative at $s=0$, this yields the result
\beq
 \int d^2p \, \ln (p\cdot p) = \int_0^\infty dp \sum_{n=1}^\infty \ln
\left( p^2 + \frac{n^2}{R^2} \right) = - \frac{\pi}{12R},
\label{5513}
\eeq
 a particular case of (\ref{558}). The double integral $d^2\xi$
and products $p\cdot \xi$ and $\xi \cdot \xi$, in the much more
complicated
expression (\ref{5511}), are to be dealt with in a similar way. Let us
introduce the basic integrals
\bea
 I_1 &\equiv& \int d^2\xi \frac{e^{ip\cdot \xi}}{\xi^2 +a^2} =
\int_0^\infty dq \sum_{m=1}^\infty \frac{e^{ipq + nm/R}}{q^2+ m^2+
a^2}= \frac{\pi}{2} \sum_{m=1}^\infty \frac{e^{inm/R} e^{-p
\sqrt{m^2+a^2}}}{\sqrt{m^2+ a^2}}, \nn \\
 I_2 &\equiv& \int d^2\xi \frac{e^{ip\cdot \xi}-1}{(\xi^2 +a^2)^2}
= \int_0^\infty dq \sum_{m=1}^\infty \frac{e^{ipq + nm/R}-1}{(q^2+
m^2+ a^2)^2} \nn \\ &=&
 \frac{\pi}{4} \sum_{m=1}^\infty \left[ \frac{e^{inm/R} e^{-p
\sqrt{m^2+a^2}}\left( 1 +p \sqrt{m^2+a^2}\right) }{(m^2+
a^2)^{3/2}} - \frac{1}{2(m^2+ a^2)^{3/2}} \right].
\label{5514}
\eea
By plugging them back into Eq. ({\ref{5511}), we obtain
\bea
V_s(R)&=& kR - \frac{(D-2)\pi}{24 R} + \frac{D-2}{2} \int dp
\sum_{n=1}^\infty \ln \left\{  4\pi e^2 \sum_{m=1}^\infty
\frac{e^{inm/R} e^{-p \sqrt{m^2+a^2}}}{\sqrt{m^2+ a^2}} \right. \label{5515}
\\ && + \left. \frac{4\pi e^2}{p^2 +n^2/R^2} \sum_{m=1}^\infty
\left[ \frac{e^{inm/R} e^{-p \sqrt{m^2+a^2}}\left( 1 +p
\sqrt{m^2+a^2}\right)-1 }{(m^2+a^2)^{3/2}} - \frac{1}{2(m^2
+a^2)^{3/2}} \right] \right\}.  \nn
\eea
No approximation is involved in Eq. (\ref{5515}). However, this
expression is quite complicated and to proceed further we need to do
some approximation, valid in the limit when $a$ is big ($a> R$). In
this case the sums can be substituted by integrals, which can then
be calculated exactly (although the results are rather long and
uninteresting and will not be written here). It is more sensible to
discuss the
first approximation in $1/a$, obtained by keeping just the leading terms
of the final result. This yields
\beq
V_s(R) \simeq kR - \frac{(D-2)\pi}{24 R} \left[ 1 - \frac{ 6\pi e^2
R^2}{a^3} \left( \gamma + 2 \ \mbox{Sinint}\, (1) - \pi  + e^{-a/R}
\right) - \frac{24 \pi e^2}{a (e^{a/R} -1 )} \right],
\label{5516}
\eeq
where $\gamma$ is Euler's constant and Sinint the standard
sinus integral (Sinint $(1)= 0.946083$). (Observe that in this
expression we have kept
a couple of terms which are representative of the asymptotically smaller
contributions to the effective potential that can be dismissed
completely in this approximation.) The dependence on $a$ and $R$ could
have been ascertained by dimensional reasons. Numerically, the result
is:
\beq
V_s(R) = kR - \frac{(D-2)\pi}{24 R} \left\{ 1 - \frac{  e^2
R^2}{a^3} \left[ 25.3416 + {\cal O}\left( \frac{1}{a} \right) \right]
\right\}.
\label{5517}
\eeq

Having at hand this result for the effective potential, we can now
study in some detail the contribution of the antisymmetric fields  to
the
static potential. From (\ref{5517}) we see immediately that the static
potential is given by expression (\ref{558}) with a  renormalized
string tension, namely
\beq
V_s(R) = k_R R - \frac{(D-2)\pi}{24 R},
\label{5518}
\eeq
where
\beq
 k_R \simeq k + 3.3172\, (D-2) \, \frac{e^2}{a^3}.
\label{5519}
\eeq
Hence, we observe that when the radius $R$ equals $R_c$, $R^2_c =
(D-2)\pi/(24 k_R)$,
it turns out that $V_s (R_c)=0$. The appearance of this critical radius,
$R_c$  indicates very probably ---as in more complicated string
models--- that
the quasi-static string picture ceases to be valid there, what has been
interpreted in Ref. \cite{7} (using a different string
model as example) as a signal for a phase transition. From
this point of view, it is now natural to interprete the effect of the
antisymmetric fields in the static potential as a renormalization of the
string constant $k$ (a one-loop correction to the classical potential),
what produces a change in the value of the critical
radius $R_c$, as compared with the one that it has in the case when
there is no coupling with antisymmetric fields.
The fact that we have been able to obtain in a quite precise way the
magnitude of this modification
 is also to be remarked.
To our knowledge, this is  the first time that such kind of
non-local calculation has been performed by using standard zeta-function
methods. And it is well known that
subleading effects, as those that we get here, can be certainly relevant
for the study
of phase transitions, because they may change in some cases the nature
of the phase transition itself.
\ms

\ni{\bf 4. \ Conclusion.}
This finishes our preliminar investigation of a string theory
coupled to  dilatonic and antisymmetric fields where, after integration
over these last ones, an effective term in the potential, of Coulomb
form, is generated. We have discussed the influence of this term
in the
static potential and we have calculated it by using the zeta-function
regularization procedure. The result can be interpreted as corresponding
to a new kind of non-local Casimir effect. Indeed, let us consider a
two-dimensional scalar field with ($D-2$)-components,
$\varphi^i$, defined in
 the space $R\times S^1$, with the following non-local Lagrangian:
\bea
L=\varphi_i (x) \left[ -\Box_x
-4e^2 \Box_x \int d^2y \,
\frac{e^{-y^\mu \partial_\mu^x}}{y^2 +a^2}
 + 8e^2 \int d^2y \, \frac{e^{-iy^\mu \partial_\mu^x}-1}{(y^2
+a^2)^2} \right] \varphi^i (x),
\label{5520}
\eea
where $i=1, \ldots, D-2$. The non-local Casimir effect in such a theory
is obtained precisely through the non-local zeta-function corresponding
to (\ref{5511}), in the same way as the previous calculation that has
been
carried out in Sect. 3. It also has the usual meaning as a quantum
correction to the free energy.

Summing up, we conclude that this study of a non-local Lagrangian
---in connection with the zeta function regularization method---
opens new possibilities to extend the fundamental concepts of
vacuum
energy or Casimir effect to completely new configurations in non-local
settings. From a different point of view, our results show clearly that
when a string is coupled to other fields, the corresponding effective
field theory, which approximately describes such a picture, might easily
be a non-local one. Finally, the remarkable power of the
zeta-function
techniques is very useful in order to deal with such ---otherwise
intractable--- situations, opening a promising new path for further
developements.
\vspace{3mm}

\noindent{\bf Acknowledgments.}
We would like to thank I.L. Buchbinder for very interesting discussions
and participation in the early stages of this work and O.Alvarez for
helpful remark.
SDO is grateful to
the members of the Department ECM, Barcelona University, for warm
hospitality.
This work has been supported by DGICYT (Spain), project
Nos. PB93-0035 and SAB93-0024, by CIRIT (Generalitat de
Catalunya), and by RFFR, project 94-020324.

\end{document}